# Plasmonic Optical Modulator based on Adiabatic Coupled Waveguides

Rui Wang[1,★], Hamed Dalir[2,★*], Xiaochuan Xu[2], Zeyu Pan[1], Shuai Sun[3], Volker J. Sorger[3], and Ray T. Chen[1,*]

[1]Department of Electrical and Computer Engineering, The University of Texas at Austin, 10100 Burnet Rd., MER 160, Austin, TX 78758, USA
[2]Omega Optics, Inc., 8500 Shoal Creek Blvd., Bldg. 4, Suite 200, Austin, TX 78757, USA
[3]Department of Electrical and Computer Engineering, George Washington University, 800 22ndSt., Science & Engineering Hall, Washington, DC 20052, USA
[★]These authors equally contributed to this work
[*]Corresponding authors: hamed.dalir@omegaoptics.com , chenrt@austin.utexas.edu

**Abstract**

In atomic multi-level systems, adiabatic elimination is a method used to minimize complicity of the system by eliminating irrelevant and strongly coupled levels by detuning them from one-another. Such a three-level system, for instance, can be mapped onto physical in form of a three-waveguide system. Actively detuning the coupling strength between the respective waveguide modes allows modulating light propagating through the device, as proposed here. The outer waveguides act as an effective two- photonic-mode system similar to ground- and excited states of a three-level atomic system, whilst the center waveguide is partially plasmonic. In adiabatic elimination regime, the amplitude of the middle waveguide oscillates much faster in comparison to the outer waveguides leading to a vanishing field build up. As a result, the middle waveguide becomes a "dark state" and hence a low insertion-loss of 8 dB is expected to keep when achieving the modulation depth as high as 70 dB despite the involvement of a plasmonic waveguide in the design presented here. The modulation mechanism relies on switching this waveguide system from a critical coupling regime to adiabatic elimination condition via electrostatically tuning the free-carrier concentration and hence the optical index of a thin ITO layer residing in the plasmonic center waveguide. This alters the effective coupling length and the phase mismatching condition thus modulation in each of outer waveguides. Our results show a modulator energy efficiency as low as 40 atto-joule per bit and an extinction ratio of 50 **dB. Given the miniscule footprint of the modulator, the resulting lumped-element limited RC delay is expected to exceed 200 GHz. Such expected performance is a direct result of both the unity-strong tunability of the plasmonic optical mode in conjunction with utilizing ultra-sensitive modal coupling between the critically-coupled and the adiabatic elimination regimes. Taken together, this new class of modulators paves the way for next-generation both energy- and speed conscience optical short-reach interconnects**.

**Main Body**

Future high-performance, multi-core, parallel computing systems require radical approaches to provide power-efficient and ultra-dense high-speed interconnects at a low-cost [**29**]. Silicon-based optical modulators is the key component in this value proposition because of its low cost, latency, and power consumption [**1,2**]. For photonic solutions to become competitive in intra- or inter-chip applications, the communications-link power consumption, including optical modulators, must be scaled down from ~10pJ/bit at present to < ~10fJ/bit in the next few years, and likely to < 1fJ/bit in the mid-term. Recently, Intel has acknowledged that the future of integrated optical circuit relies on technologies that aggressively reduce power consumption with reasonable performance [**3**]. As of today, two types of optical modulators (refractive- or absorptive-based) are available. For the refractive approach, changing the plasma dispersion effect or free carrier absorption in silicon manipulates the real part of material permittivity and hence the modulation occurs. Nevertheless, due to a small change in the real part of material permittivity modulators such as Mach–Zehnder's demand several hundreds of micrometers to acquire sufficient extinction ratios required to achieve the desired bit-error-rate of the signal. This results in a large footprint, low operation speed and eventually large power consumption modulators [**4,5**]. To overcome these limits, other refractive modulators operated with multiple-pass but in single-beam interference, such as disk or ring resonator-based devices. Due to the nature of the resonance, a narrowband modulation of 0.1 nm with a stringent fabrication process is expected [**6,7**]. In addition, a precise temperature stabilization to hold the device on resonance is indispensable, requiring a high static power consumption [**8**]. On the contrary, absorptive-based modulators manipulate the transmission intensity directly, that is without the need for interference, by means of modifying the imaginary part of the material permittivity such as saturable absorption, electro-absorption, the Franz–Keldysh effect (FKE) and the quantum-confined Stark effect (QCSE) [**9-11**]. Germanium modulators (FKE-based) offer a relatively high-speed operation but suffer from a large insertion loss related to the intrinsic GeSi absorption loss at zero bias voltage [**9-11**]. Furthermore, CMOS-compatible integration comes with stringent process requirements such as found in epitaxial growth, wafer- or die- bonding, bonding, which limits the liability of the final devices [**9-11**]. Recently, optical modulators based on two-dimensional (2D) materials such as graphene devices provide advantages of compact footprint and ultrafast modulation speed across a broad range of wavelengths [**8,12,13**]. Nevertheless, due to the poor overlap of the light with the thin 2D absorber, relatively large biasing voltage is required to attain modulation [**13**]. Previously, we have experimentally demonstrated a novel plasmonic modulator based on a thin layer of Indium thin oxide (ITO) with 1 dB per micrometer modulation strength, while showing improvement over graphene devices, this concept still requires either large power consumption or high insertion loss for real optical interconnect applications (> 10 dB) [**14**].

In this work, we propose an atto-joule, low-loss; high-speed plasmonic optical modulator based on tunability of three adiabatically coupled waveguides (ACW). Analogous to adiabatic elimination (AE) in an atomic three-level system, two outer waveguides function as ground- and excited- states, where the intermediate waveguide is equivalent to the "dark state" (effective two-level system). In atomic three-level systems, adiabatic elimination is a method that used to minimize complicity of the system by eliminating irrelevant, strongly coupled nearby levels with a very large detuning between them. In this case, amplitude of the middle level oscillates much faster in comparison to the other levels leading in no significant build up. As a result, the three-level system reduces to an effective two-level system with a vanishing coupling strength (i.e. diverging mode coupling length) between the ground- and excited- states, with the intermediate level become 'dark' state [15]. The dark state, namely the middle waveguide has a plasmonic metal-oxide-semiconductor (MOS) configuration, which concentrates the optical modes' transverse magnetic (TM) field into a thin ITO layer, thus allowing for a high modal overlap enabling efficient tunability [16-18]. The modulation mechanism originates from electrically changing the free carrier concentration of the ITO layer that actively controls the effective coupling of the outer waveguides; consequently, the coupling strength and phase mismatching condition change and modulation take place. Our simulation results indicate that, the proposed device will operate with a power consumption as low as 40 atto-joule and 50 dB modulation depth. In addition, our ACW plasmonic modulator provides high-speed operation exceeding 250 GHz and insertion loss of 8 dB at the modulation depth of 70 dB. The proposed device can enable novel architectures for on-chip optical communications.

Our compact plasmonic modulator is comprised of three-coupled waveguides, where the outer waveguides are two identical Silicon photonics strip waveguides (Fig. 1). Intermediate waveguide consists of a metallic Au-pad with a gate oxide of 20 nm-thick $SiO_2$ and an active ITO with the thickness of 10 nm sandwiched between them.  Here coupling between the two outer waveguides becomes controllable through varying a bias voltage across the MOS capacitor in intermediate waveguide. This leads to the mode's index change altering the propagation mismatch and hence the inversion length of the outer waveguides. The three coupled mode equations for this symmetric structure is written in Eq. (1) with $\varDelta\beta = \beta_1 - \beta_2$, where $\beta_1$ and $\beta_2$ are the propagation constants of the outer and middle waveguides, respectively. And $\kappa_{13}$ is the coupling strength between the identical outer waveguides. Also $\kappa_{12}$ are $\kappa_{21}$ are the coupling strengths between the outer and middle waveguides [28]. Due to the symmetry, $\kappa_{12} = \kappa_{32}$ and $\kappa_{21} = \kappa_{23}$, when $\kappa_{21} \ll |\varDelta\beta|$, the AE condition is realized [15] and the effective coupling strength between the two outer waveguides is $\kappa_{eff} = \kappa_{13} + \kappa_{12}\kappa_{23}/\varDelta\beta$ [15]. As shown in figure 1(b), the TM mode injected through one of the outer waveguides propagates merely in the other outer waveguide. This confirms the AE condition, where the center plasmonic waveguide is in dark state. The outer waveguides are identical with a width of 250 nm, whereas the center waveguide is 110 nm wide, the total gap between the outer waveguides is 150 nm (i.e. ~λ/10) with the height of 550 nm-thick silicon, while the general platform is silicon-on-insulator (SOI).

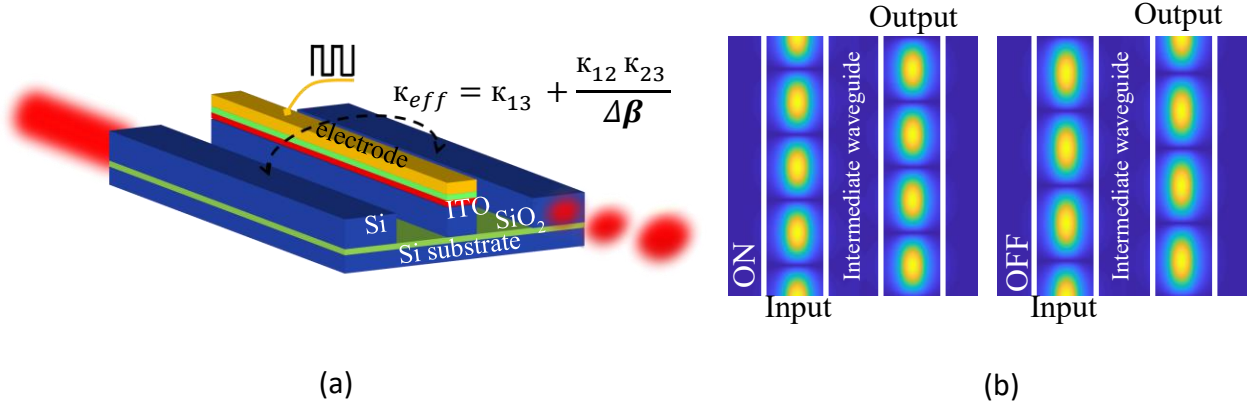

**Fig. 1.** The (a) Schematic structure and (b) fundamental TM mode of the ACW waveguides. The corresponding effective indices at 1550 nm are 2.505-0i and 1.878-0.0105i for ON- and OFF- states, respectively. Outer waveguides are identical with the width of 250 nm and middle waveguide is 110 nm, spacing of the outer waveguides to middle waveguides is 20 nm, the height of the waveguides is 550 nm.

$$\frac{d}{dz}\begin{vmatrix}A_1\\A_2\\A_3\end{vmatrix} = j\begin{vmatrix}0 & \kappa_{12}exp(-j\Delta\beta_{12}z) & \kappa_{13}\\ \kappa_{21}exp(j\Delta\beta_{12}z) & 0 & \kappa_{23}exp(j\Delta\beta_{12}z)\\ \kappa_{13} & \kappa_{32}exp(-j\Delta\beta_{12}z) & 0\end{vmatrix} \cdot \begin{vmatrix}A_1\\A_2\\A_3\end{vmatrix} \quad (1)$$

Previously we experimentally reported that the N-type doping concentration of an ITO at zero biasing voltage is 1e19 cm$^{-3}$ under certain process conditions [**19-21**]. Accordingly, the accumulated ITO electron concentration changes to 5e20 cm$^{-3}$ when the biasing voltage raises to two volts. Thus, the accumulated electron concentration in ITO is $5e20\ cm^{-3}\ V_d/2V$. As shown in figure 1 (b), even a sub-volt operation will completely turn the ON- to OFF- state for a device with tens of micrometers-long.

Figure 2 (a) and (b) show the modulation depth and insertion loss, respectively, based on the device length under various bias voltages. Here the first and second peaks of the modulation depth for ACW are corresponding to the ON- ($V_d$= 0 V) and OFF-states, respectively. As illustrated in figure 2 (a) the second peak, which is associated with the minimum of the optical power at 0 V (ON state), will remain the same. On the other hand, the peak associated with the minimum of the optical power in the OFF state varies as voltage changes. While increasing the biasing voltage increases the absorption and consequently, the optimized modulation length corresponding to the first peak decreases.

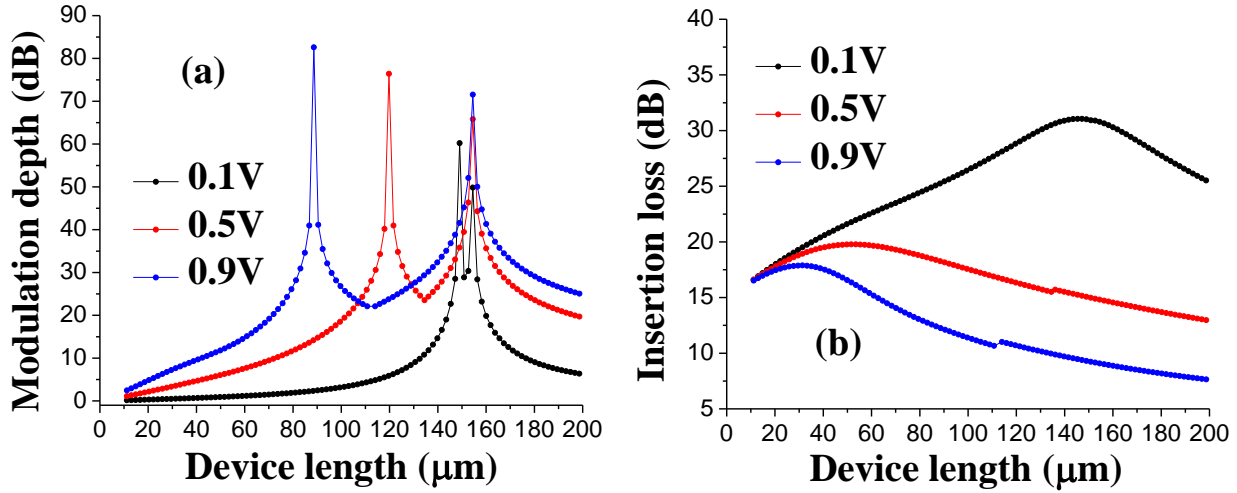

**Fig.2.** The evolution of the (a) modulation depth and (b) insertion loss of the conventional and ACW ITO with the device length at different bias voltage. The wavelength is 1.55μm.

As shown in figure 2 (b) optical absorption in the center waveguide enlarges with increased device length and applied bias voltage. However due to the phase detuning of the ON state from $(2N+1)\pi$ in the output port, the overlap of optical power with the center waveguide reduces with the longer device length and higher bias voltage. As a result, insertion loss for longer device length become smaller. The kinks in the insertion loss - device length profiles are due to the switching from ON to OFF states. At the sampled device lengths before and after the kinks, the insertion loss is associated with the optical power at $V_d = 0$ V and $V_d > 0$ V, respectively. In figure 3, we compare the modulation depth and insertion loss of our ACW modulator against conventional device designs. Here, "conventional A" and "conventional B" denote conventional plasmonics waveguides with the silicon layer of 110 nm and 650 nm wide, respectively (height is fixed at 550 nm). A metallic Au-pad with a gate oxide of 20 nm-thick $SiO_2$ and an active ITO with the thickness of 10 nm sandwiched between them has covered the top surface of the waveguides. As shown in figure 3 (a) our proposed ACW device offers higher modulation depth exceeding 50 dB with bias voltages as low as 100 mV. Figure 3(c) shows the device footprint for our ACW and conventional A and conventional B modulators. As shown in figure 3 (b) and (c) for conventional ITO modulators, there is a trade of between insertion loss and footprint of the device. For instance, reduction of insertion loss at sub-volt operation with modulation depth of 30 dB, requires shrinkage in the width of waveguide and consequently increase of the length of modulator. In the contrary, our ACW modulator exhibits a low insertion loss of 8 dB for 0.9 V biased voltage with much smaller device length of 155µm.

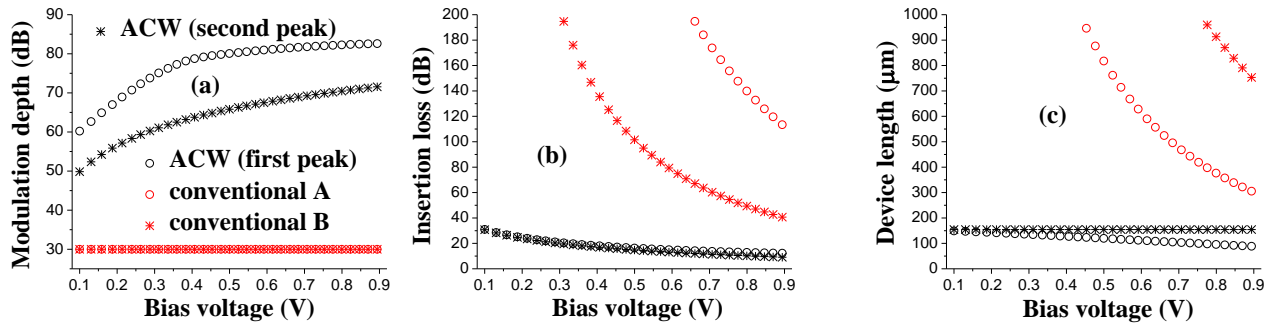

**Fig.3.** Comparison of the (a) modulation depth, (b) insertion loss and (c) device length of the conventional ITO plasmonic- and our ACW- modulator. Here "conventional A" and "conventional B" denote to a plasmonic single silicon waveguide with 550 nm height and width of 110 nm and 650 nm, respectively. A metallic Au-pad with a gate oxide of 20 nm-thick $SiO_2$ and an active ITO with the thickness of 10 nm sandwiched between them has covered the top surface of the waveguides. The legend also applies to Fig.4.

Next, we calculated power consumption (P=0.25$C_{ox}V_d^2$) and the lumped-element 3-dB bandwidth ($f_{3dB}$=1/2π$RC_{ox}$). Here $C_{ox} = C_{ox,unit}WL$ where $C_{ox,unit}$ is the unit-area MOS capacitance, L is the device length and W is the width of the ITO layer. R is the resistance in the ITO layer and is expressed as $R = \rho_{ITO}W/(2t_{ITO}L)$ where $\rho_{ITO}$ and $t_{ITO}$ are the resistivity and thickness of the ITO. $\rho_{ITO} = 1/(n_{ITO}q\mu_{ITO})$ where $n_{ITO}$ and $\mu_{ITO}$ are the ITO doping density and electron mobility, respectively, with $\mu_{ITO} \approx 100\ cm^2/Vs$ [27]. Note that the lumped-element 3-dB bandwidth depends on W only. At W=110nm and W=650nm, the lumped-element 3-dB bandwidths are 4.5 THz and 130 GHz, respectively. As shown in figure 4 (a) our ACW modulator operates <5 fJ, while the conventional ITO modulator with reasonable loss, requires tens of femto-joule. In addition, we obtained a power consumption of 40 atto-joule corresponding to a driving voltage of 0.1 V in figure 4 (a). The modulation response of >200 GHz is expected from the second peak in figure 4 (b), while the insertion loss can be 8 dB at the device length of 155 µm for sub-volt biasing in figure 3 (b), further reduction on the insertion loss is expected with optimization of the structure as well as increase of the biased voltage.

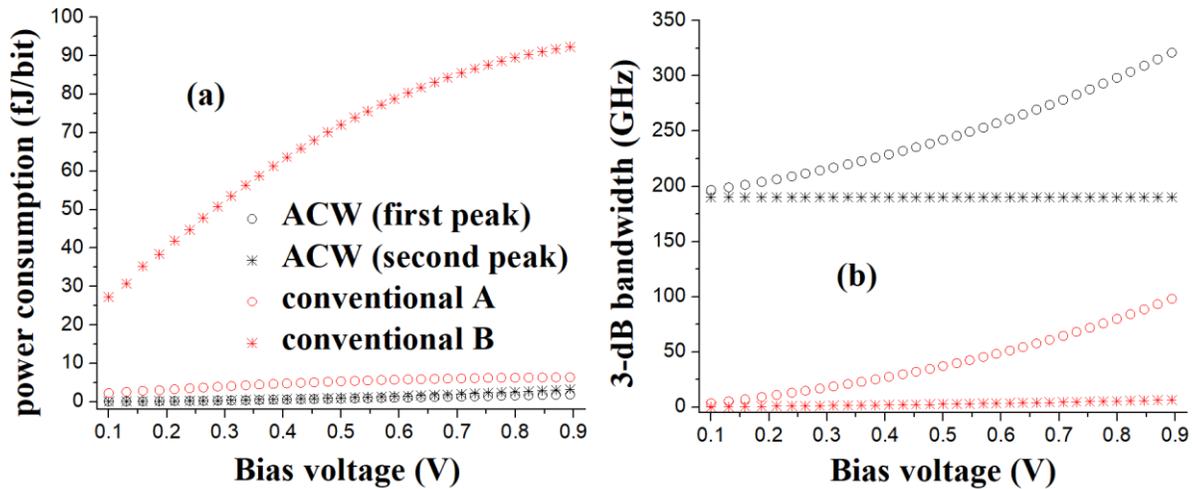

**Fig.4.** The evolutions of (a) power consumption and (b) 3-dB bandwidth with the bias voltage at the device lengths corresponding to the first and second peaks. The wavelength is 1.55$\mu$m.

Finally, we set the device length to 90 $\mu m$ corresponding to the first peak at $V_d = 0.9$ V. As shown in figure 5, we studied the spectral response of the modulator via swiping wavelength of the injected light. The results show a switch of the light propagating from one of the outer waveguide to another. An increase of the biased voltage will enhance the broadband functionalities of our propose device. Details will be discussed elsewhere.

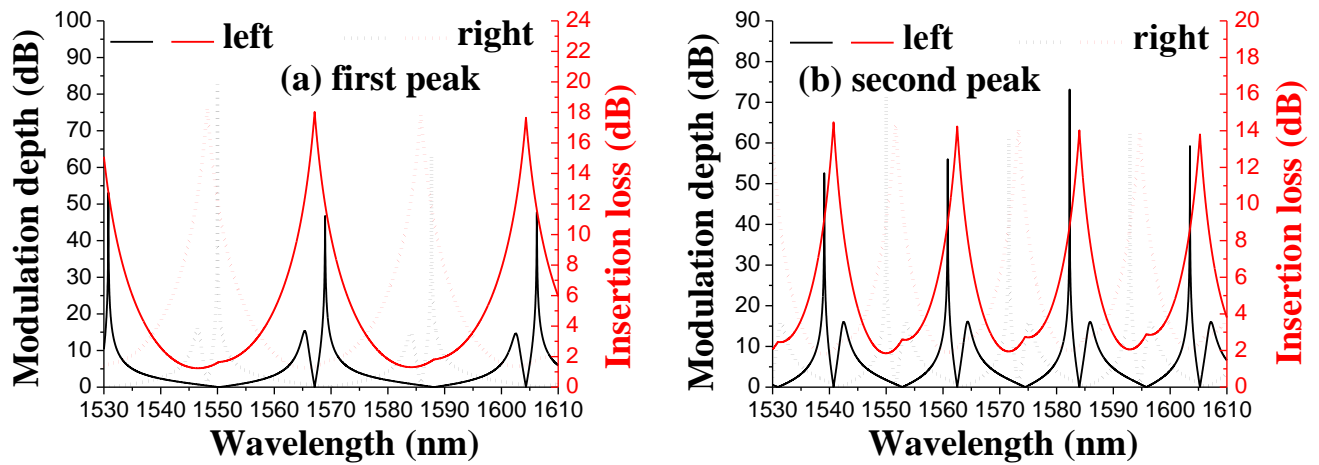

**Fig.5.** The spectrums of the modulation depth and insertion loss in the left and right arms when the ACW is at the (a) first and (b) second peaks of the modulation depth. Device length is 90 µm at the bias voltage of 0.9V.

In conclusion, we have developed a new concept for an energy-efficient, high-speed and low-loss ITO plasmonic modulator. The underlying operation principle is based on altering the coupling strength of an adiabatically coupled three-waveguide system. By voltage-tuning a plasmonic mode via carrier modulation of an active ITO layer of the center waveguide of the modulator, we show that tunability in and out-of an adiabatic elimination region in the index-design space enables dramatic modulation depth of exceeding 50 dB for just 100 mV of applied bias. The resulting 40 atto-Joule per bit efficiency opens a new era for electro-optic modulation in Silicon photonics. With the design scheme for the device length balancing the modulation depth and insertion loss by maximizing their product, the ACW modulator obtains superior performances over the current modulator techniques. The proposed modulator exemplifies the recently discussed device performance [**22**] paving the way for the atto-joule efficient, fast and low-power-consumption on-chip optoelectronic devices [**23-25**] enabling new functionality for photonics interconnects [**26**].

**Author contributions**

H.D. and V.J.S. conceived the idea. H.D. and R.T.C. supervised this project. H.D. and R.W. performed the theoretical analysis. R.W. performed the numerical simulations. H.D. and R.W. wrote the manuscript. All authors discussed the manuscript.

**Acknowledgement**

We are grateful for the support of Air Force Office of Scientific Research (AFOSR) Small Business Innovation Research (SBIR) program under award number FA9550-17-P-0014